\documentclass[aps,twocolumn,prd,dvipsnames,usenames,floatfix,superscriptaddress,nofootinbib]{revtex4-1}
\pdfoutput=1

\usepackage{footnote}
\makesavenoteenv{tabular}
\makesavenoteenv{table}
\usepackage[utf8x]{inputenc}
\usepackage{units}
\usepackage[english]{babel}
\usepackage{xspace}
\usepackage{paralist}
\usepackage{amsmath} 
\usepackage{amssymb} 
\usepackage{graphicx}
\usepackage{bm}
\usepackage{tabularx}
\usepackage[final,textsize=footnotesize,color=ForestGreen!05]{todonotes}
\PassOptionsToPackage{unicode}{hyperref}
\PassOptionsToPackage{bookmarks=false}{hyperref}
\usepackage{hyperref}

\newcommand\myshade{85}
\colorlet{mylinkcolor}{violet}
\colorlet{mycitecolor}{Aquamarine}
\colorlet{myurlcolor}{YellowOrange}

\hypersetup{
 linkcolor  = mylinkcolor,
 citecolor  = mycitecolor!\myshade!black,
 urlcolor   = myurlcolor!\myshade!black,
colorlinks = true,
}

\newcommand{\myparagraph}[1]{\bigskip\noindent\textbf{#1}}
\newcommand{\Br}{\mathrm{Br}}
\newcommand{\decay}{\mathrm{dec}}
\newcommand{\events}{\mathrm{events}}
\renewcommand{\min}{\mathrm{min}}
\renewcommand{\max}{\mathrm{max}}
\begin{document}

\title{Probing baryon asymmetry of the Universe at LHC and SHiP}

\author{Iryna~Boiarska}
\affiliation{Discovery Center, Niels Bohr Institute, Copenhagen University, Blegdamsvej 17, DK-2100 Copenhagen, Denmark} 
\author{Kyrylo~Bondarenko}
\author{Alexey~Boyarsky}
\author{Shintaro~Eijima}
\author{Maksym~Ovchynnikov}
\affiliation{Intituut-Lorentz, Leiden University, Niels Bohrweg 2, 2333 CA Leiden, The Netherlands} 
\author{Oleg~Ruchayskiy}
\affiliation{Discovery Center, Niels Bohr Institute, Copenhagen University, Blegdamsvej 17, DK-2100 Copenhagen, Denmark} 
\author{Inar Timiryasov}
\affiliation{Institute of Physics, Laboratory for Particle Physics and Cosmology,
\'Ecole Polytechnique F\'ed\'erale de Lausanne, CH-1015 Lausanne, Switzerland}

\begin{abstract}
  The origin of the baryon asymmetry of the Universe (BAU) is one of the major puzzles beyond the Standard Model. Detecting the particles responsible for the generation of the BAU  would  enormously boost our understanding of physics of the early Universe. Here we demonstrate that searches for displaced vertices at  ATLAS, CMS and LHCb allow detecting heavy neutral leptons (HNL) with parameters that can simultaneously lead to the successful generation of the BAU and explain the masses and oscillations of active neutrinos.
 The combination of a dedicated LHC search program and a complementary  Intensity Frontier experiment (such as SHiP)   will allow exploring a sizeable part of the ``minimal HNL baryogenesis'' parameter space.
\end{abstract}
\maketitle 

\myparagraph{Introduction.}
Heavy Neutral Leptons (also known as HNLs, right-handed, Majorana or sterile neutrinos) can provide resolutions to several beyond-the-Standard-Model puzzles.
They can explain neutrino masses and oscillations (via the so-called type I seesaw mechanism~\cite{Minkowski:1977sc,Yanagida:1979as,Glashow:1979nm,GellMann:1980vs,Mohapatra:1979ia,Mohapatra:1980yp}); can generate the baryon asymmetry of the Universe via the process known as \emph{leptogenesis} (see reviews~\cite{Buchmuller:2004nz,Davidson:2008bu,Shaposhnikov:2009zzb,Pilaftsis:2009pk,Drewes:2017zyw} and references therein); and can provide a dark matter candidate (see e.g.\ \cite{Boyarsky:2018tvu} for review).
Moreover, all three phenomena can be explained within one compact extension of the Standard Model, known as the \emph{Neutrino Minimal Standard Model} or $\nu$MSM~\cite{Asaka:2005pn,Asaka:2005an}, see~\cite{Boyarsky:2009ix} for review.
In particular, the mass scale of the HNLs responsible for the generation of BAU can be as low as GeV~\cite{Akhmedov:1998qx,Asaka:2005pn,Asaka:2005an}, thus opening the possibility of probing a leptogenesis scenario in particle physics laboratories.
Many subsequent works have investigated the leptogenesis with GeV-scale HNLs, see e.g.~\cite{Shaposhnikov:2008pf,Canetti:2010aw,Asaka:2010kk,Anisimov:2010gy,Asaka:2011wq,Besak:2012qm,Canetti:2012vf,Drewes:2012ma,Canetti:2012kh,Shuve:2014zua,Bodeker:2014hqa,Abada:2015rta,Hernandez:2015wna,Ghiglieri:2016xye,Hambye:2016sby,Drewes:2016lqo,Asaka:2016zib,Drewes:2016gmt,Hernandez:2016kel,Drewes:2016jae,Asaka:2017rdj,Eijima:2017anv,Ghiglieri:2017gjz,Eijima:2017cxr,Ghiglieri:2017csp,Eijima:2018qke}.
At the same time, searches for heavy neutral leptons have been performed and are included into the scientific plans of most of the currently running particle physics experiments~\cite{Chatrchyan:2012fla,Aaij:2014aba,Aad:2015xaa,Khachatryan:2015gha,Khachatryan:2016olu,CortinaGil:2017mqf,Mermod:2017ceo,Izmaylov:2017lkv,Sirunyan:2018mtv}.
Many search strategies have been proposed, some of them having potential to reach deep into the HNL's parameter space using the \textit{high luminosity} phase of the LHC~\cite{Helo:2010cw,Liventsev:2013zz,Abada:2013aba,Helo:2013esa,Canetti:2014dka,Gago:2015vma,Das:2015toa,Banerjee:2015gca,Izaguirre:2015pga,Antusch:2015mia,Arganda:2015ija,Degrande:2016aje,Antusch:2017pkq,Ruiz:2017yyf,Antusch:2017hhu,Dube:2017jgo,Cai:2017mow,Antusch:2018ahh,Deppisch:2018eth,Abada:2018sfh,Cottin:2018nms,Marcano:2018fto,Drewes:2018gkc,Dib:2018iyr}.
A number of proposed experiments will further probe the HNL's parameter space~\cite{Anelli:2015pba,Alekhin:2015byh,Gligorov:2017nwh,Curtin:2018mvb,Feng:2017uoz,Kling:2018wct, Baer:2013cma,Brau:2015ppa,CEPC-SPPCStudyGroup:2015csa,Gomez-Ceballos:2013zzn}.

The question whether the HNLs, probed in particle physics experiments, can be responsible for the generation of the BAU has been addressed before (see e.g.~\cite{Antusch:2017pkq} for review).
The current work is motivated by several recent theoretical and experimental developments:
\begin{asparaenum}[(1)]
\item The recent work~\cite{Eijima:2018qke} has elaborated the region in the parameter space where successful baryogenesis is possible.
  In particular it has been demonstrated that HNLs with \emph{larger than
    previously estimated} mixing angles can lead to successful
  baryogenesis.
  Ref.~\cite{Eijima:2018qke} has been limited to HNLs with masses $m_{N} \lesssim \unit[10]{GeV}$.
  \emph{In this work we extend the results
    of~\cite{Eijima:2018qke} to higher masses.}

\item The HNL searches with displaced vertices (DV) at ATLAS and CMS experiments have been discussed in~\cite{Cottin:2018nms}.
  \emph{We revise this strategy and discuss another DV strategy for CMS.}

\item The estimates of the sensitivity of the LHCb experiment have
  concentrated on HNLs produced from the $W$ bosons~\cite{Antusch:2017hhu}.
  Searches~\cite{Aaij:2014aba} concentrated on a rare production channel $B \to \mu N$ followed by $ N \to \mu \pi$. \emph{We update the sensitivity estimates by including the HNLs produced in the decays of the $B$  mesons and decaying to different final states.}

\item The sensitivity of the SHiP experiment for heavy neutral leptons has been revised~\cite{SHiP:2018xqw}, in particular for masses $m_N \sim \unit[5]{GeV}$ and mixing angles\footnote{Flavor dependent mixing angles
$U_e^2, U_\mu^2, U_\tau^2$ specify by how much the interaction of HNLs with $W/Z$ bosons is \emph{weaker} than that of the Standard Model neutrinos. The total mixing  is defined as $U^2 = \sum_\alpha |U_\alpha|^2$.} $U^2 \gtrsim 10^{-8}$.
\emph{We show that the synergy between the LHC  and the intensity frontier searches can allow covering a sizeable part of the parameter space where successful baryogenesis can take place. }
\end{asparaenum}

The paper is organized as follows.  First we summarize recent baryogenesis results and extend them to higher masses. 
Then we give a brief overview of the DV search strategies at the LHC.
Sec.~\ref{sec:sensitivity-estimates} presents our estimates of the sensitivity in the DV searches at ATLAS, CMS and LHCb.
In Sec.~\ref{sec:results} we present our results.
\begin{table*}[t]
  \centering
  \begin{tabular}{l|c|c|c|c|c}
    Parent/Experiment & $l_\min, \ l_\max$ & Cross-section & Number & $\langle E\rangle$  \\
    \hline $W$ @ ATLAS/CMS, Short DV & \unit[0.4]{cm}, \unit[30]{cm}~\cite{Cottin:2018nms} & $\sigma_W \simeq \unit[193]{nb}$~\cite{Aad:2016naf} & $5\times 10^{11}$ &
    -- \\
    $W$ @ CMS, Long DV & \unit[2]{cm}, \unit[300]{cm} & $\sigma_W \simeq \unit[193]{nb}$~\cite{Aad:2016naf} &$5\times 10^{11}$ & -- \\
    $B$  @ LHCb & \unit[2]{cm}, \unit[60]{cm}~\cite{Antusch:2017hhu} &$\sigma_{b\bar b} \simeq \unit[1.3\times 10^8]{pb}$~\cite{LHCb-bbbar} & $4.9\cdot 10^{13}$ & $\unit[84]{GeV}$~\cite{Cacciari:1998it}\\
  \end{tabular}
  \caption[Number of parent particles]{Number of parent particles ($W$ bosons or $B$  mesons), their average energy $\langle E\rangle$ and corresponding parameters of the DV search schemes used for the estimates. All numbers are given for the high luminosity phase of the LHC ($\mathcal{L} = \unit[3000]{fb^{-1}}$ for CMS/ATLAS and $\mathcal{L} = 380\text{ fb}^{-1}$ for the LHCb) and can be proportionally rescaled for other luminosities. To obtain the average energy of $B$ mesons we use their spectrum  at $\sqrt{s} = 13\text{ TeV}$ in the pseudorapidity range of LHCb as provided by FONLL simulations~\cite{Cacciari:1998it}.}
  \label{tab:parent}
\end{table*}

\myparagraph{Leptogenesis with GeV-scale HNLs.}
\label{sec:leptogenesis}
At least two HNLs, highly degenerate in mass  are needed for the successful
baryogenesis in the GeV-mass range~\cite{Asaka:2005an,Asaka:2005pn}. 
Ref.~\cite{Eijima:2018qke} has focused on the region of the parameter space $\unit[0.1]{GeV} \lesssim m_N \lesssim \unit[10]{GeV}$, which is important for the intensity frontier experiments.
Since the DV signature at the LHC allows probing heavier masses, we extent
the analysis of~\cite{Eijima:2018qke} by computing the rates entering the
kinetic equations describing the generation of baryon asymmetry for masses
$\unit[10]{GeV} \lesssim m_N \lesssim \unit[30]{GeV}$, and perform a scan over the parameter space. Even heavier HNLs require yet another dedicated study. Indeed, as suggested in~\cite{Blondel:2014bra}, the decays $N\to W \ell$---when kinematically allowed---lead to efficient washout of the generated asymmetry. Therefore, one can expect that the allowed region is bounded by $m_N\lesssim M_W$. This effect is not included in our kinetic equations and deserves a dedicated study.

Note that in the $\nu$MSM the masses of the HNLs must be nearly degenerate since their oscillations play the crucial role in generating of the BAU. Therefore, two Majorana HNLs could be effectively described by a single Dirac particle \cite{Shaposhnikov:2006nn} with the corrections of order $\Delta M/M$, where $\Delta M$ is the mass splitting. This means that the amplitudes of the lepton number violating processes (including same-sign lepton signatures) are suppressed.

\myparagraph{Displaced vertices at the LHC.}
\label{sec:dv-at-lhc}
Current experimental bounds (summarized e.g.
in~\cite{Alekhin:2015byh}) put the mixing of HNLs at the level $U^2 \sim 10^{-5}$ for masses $m_N \gtrsim \unit[2]{GeV}$.
HNLs with such parameters would travel distances of the order of meters for $\gamma$ factor $\sim 1$~\cite{Bondarenko:2018ptm}.
Owing to a sufficiently long lifetime of HNLs one can search for their decays at macroscopic distance from the interaction point -- displaced vertex (DV) signature.
Such a search strategy drastically reduces possible backgrounds~\cite{Antusch:2016vyf,Abada:2018sfh,Antusch:2017hhu,Gago:2015vma,Helo:2013esa,Cottin:2018nms}, making this possibility particularly interesting at colliders.
Usually the DV searches are performed using the inner trackers of ATLAS or CMS. 
In~\cite{Bondarenko:2019} it was proposed to utilize CMS muon tracker to reconstruct the HNLs that decay into pairs of muons. We call the latter strategy ``\emph{Long DV}'' as this schema allows probing
longer displacements. Correspondingly the other strategy will be called ``\emph{short DV}''.

\begin{figure}[!h]
  \centering
  \includegraphics[width=\linewidth]{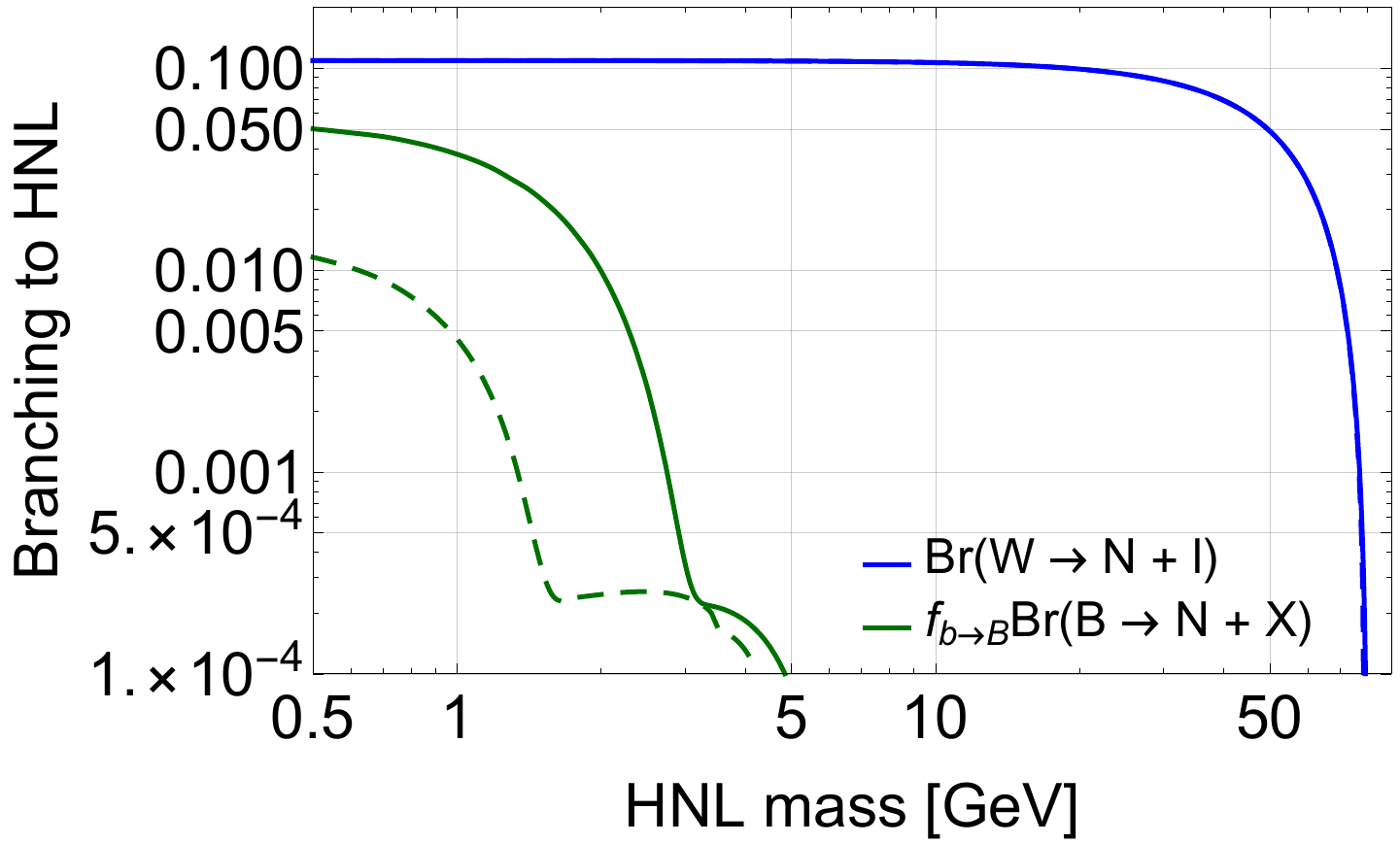}
  \caption{Branching fraction of the production of the HNL from $B$  mesons and $W$ bosons for the mixing angles $U_\alpha ^2 = 1$. For the $B$  mesons the solid line shows the branching ratio for the mixing with $\nu_e$ or $\nu_\mu$, while the dashed line corresponds to the mixing with $\nu_{\tau}$. The fragmentation fractions $f_{b \to B}$ (the probability for a $b$ quark to hadronize into a specific meson) is taken from~\cite{Aaij:2011jp,Aaij:2017kea,Kling:2018wct}.}
  \label{fig:branching}
\end{figure}

\section{Sensitivity estimates}
\label{sec:sensitivity-estimates}
The number of detected events is given by
\begin{equation}
  \label{eq:n-events-w}
  N_{\events} = N_{\rm parent} \cdot \Br\cdot P_{\decay}\cdot \epsilon,
\end{equation}
where $N_{\rm parent}$ is the number of parent particles ($W$ bosons and/or
$B$ mesons)\footnote{Production from $Z$ and/or Higgs bosons, as well as direct
  production via Drell-Yan processes and quark/gluon fusion, is
  subdominant~\cite{Gago:2015vma,Degrande:2016aje,Das:2017rsu,Ruiz:2017yyf,Abada:2018sfh} and will be
  neglected in what follows.} shown in Table~\ref{tab:parent}; $\Br$ is the
branching fraction for $W^\pm \to N + \ell^\pm$ or (semi-)leptonic decays of $B$ mesons, see Fig.~\ref{fig:branching};
$P_{\decay}$ is the decay probability
 \begin{equation}
   P_{\decay} = e^{-l_{\min}/c\tau \langle \gamma_{N}\rangle}-e^{-l_{\max}/c\tau \langle \gamma_{N}\rangle}
   \label{eq:Pdecay}
\end{equation}
with $l_\min$ and $l_\max$ determined by the geometry of a tracker, and
$\langle \gamma_N\rangle$ being the average $\gamma$ factor of the
HNL.\footnote{For the HNLs produced from $B$ mesons we define $ \langle \gamma_{N}\rangle = \sqrt{\langle \gamma_{X}\rangle^{2} (\gamma^{\text{rest}}_{N})^{2}-1},$ where $\gamma_X$ is the $\gamma$ factor of a parent meson that moves in the direction of the decay volume, and $\gamma^{\text{rest}}_{N}$ is the $\gamma$ factor of the HNL in the rest frame of the meson. 
For the HNLs from $W$ bosons we ran MadGraph5~\cite{Alwall:2014hca}
simulation to find the average $\gamma$ factor.}
Finally, the parameter $\epsilon$ is the \emph{efficiency}---the fraction of all HNL decays that occurred inside the decay volume (between $l_\min$ and $l_\max$) that have passed the selection criteria and were successfully reconstructed.

  \bigskip
  
The sensitivity curve is determined by the condition $N_\events \approx 3$ (95\% confidence limit). The lower boundary of such a curve is determined by the regime $l_{\min} \ll c\tau_{N}\langle \gamma_{N}\rangle$ where the number of events scales approximately as $N_\events \propto U^4$ which allows to find the $U^2$ value for the lower boundary from Eqs.~\eqref{eq:n-events-w}--\eqref{eq:Pdecay}.
The upper boundary is found in the regime $l_{\text{min}}>c\tau_{N}\langle \gamma_{N}\rangle$ so that $P_{\decay} \approx \exp\left[-l_{\min}/(c\tau \gamma_{N})\right]$ and $N_{\text{events}}$ is exponentially sensitive to $\gamma_{N}$. Therefore the HNL momentum distribution becomes important with the most energetic HNLs determining the exact shape of the boundary. By assuming that all of the HNLs are produced with the average $\gamma$ factor, we \emph{underestimate} the position of the upper boundary and, as a result, the maximal mass probed, which is defined as the intersection of the lower and the upper bounds of the sensitivity.

\myparagraph{Short DV at ATLAS and CMS.}
The analysis of this case has been performed in~\cite{Cottin:2018nms} using
Monte Carlo (MC) simulations. 
Below we compare their result with our analytic estimates.

The efficiency of the HNL detection is determined by the cuts imposed in order to reduce background and efficiently reconstruct displaced vertex in the ATLAS inner detector~\cite{Cottin:2018kmq}:
\begin{compactitem}[--]
    \item One prompt lepton ($e$ or $\mu$) with $p_{T} >\unit[25]{GeV}$ that  serves for tagging of the process $W \to \ell+N$. In case of mixing with $\tau$, the reconstruction of prompt $\tau$ leads to the reduced efficiency.
    \item The distance between the interaction point and the decay position must be between $l_{\min}$ and $l_{\max}$.
    \item There should be at least four charged decay products with $p_{T}>1\text{ GeV}$ and transverse impact parameter $|d_{0}| > 2\text{ mm}$.
    \item The invariant mass of the DV reconstructed by the selected decay products must be larger than $5\text{ GeV}$.
\end{compactitem}
The last two criteria reduce the search to the background free region (see a discussion in~\cite{Cottin:2018nms}).

In our estimates we use efficiencies $\epsilon$ provided by the authors of~\cite{Cottin:2018nms}.\footnote{We are grateful to the authors of~\cite{Cottin:2018nms} for sharing with us the results of their MC simulations.}
To obtain the average energy of the HNLs $\langle E_{N}\rangle$ and the geometric acceptance $\epsilon_{\text{geom}}$\footnote{We define the
geometrical acceptance as the amount of the HNLs that fly in the direction of ATLAS experiment.} we simulated  the process $p p \to W +X$ to leading order in MadGraph 5~\cite{Alwall:2014hca}.
Using the resulting spectrum of the $W$ bosons, we calculated the energy distribution of the HNLs in the pseudorapidity range $|\eta_{N}| < 2.5$ (see Appendix~\ref{sec:production} for details). In the mass range $m_{N}\lesssim 30\text{ GeV}$ we obtained $\langle E_{N}\rangle \approx 80\text{ GeV}$ and $\epsilon_{\text{geom}} \approx 0.5$.

\begin{figure}[t!]
    \centering
    \includegraphics[width=0.45\textwidth,draft=false]{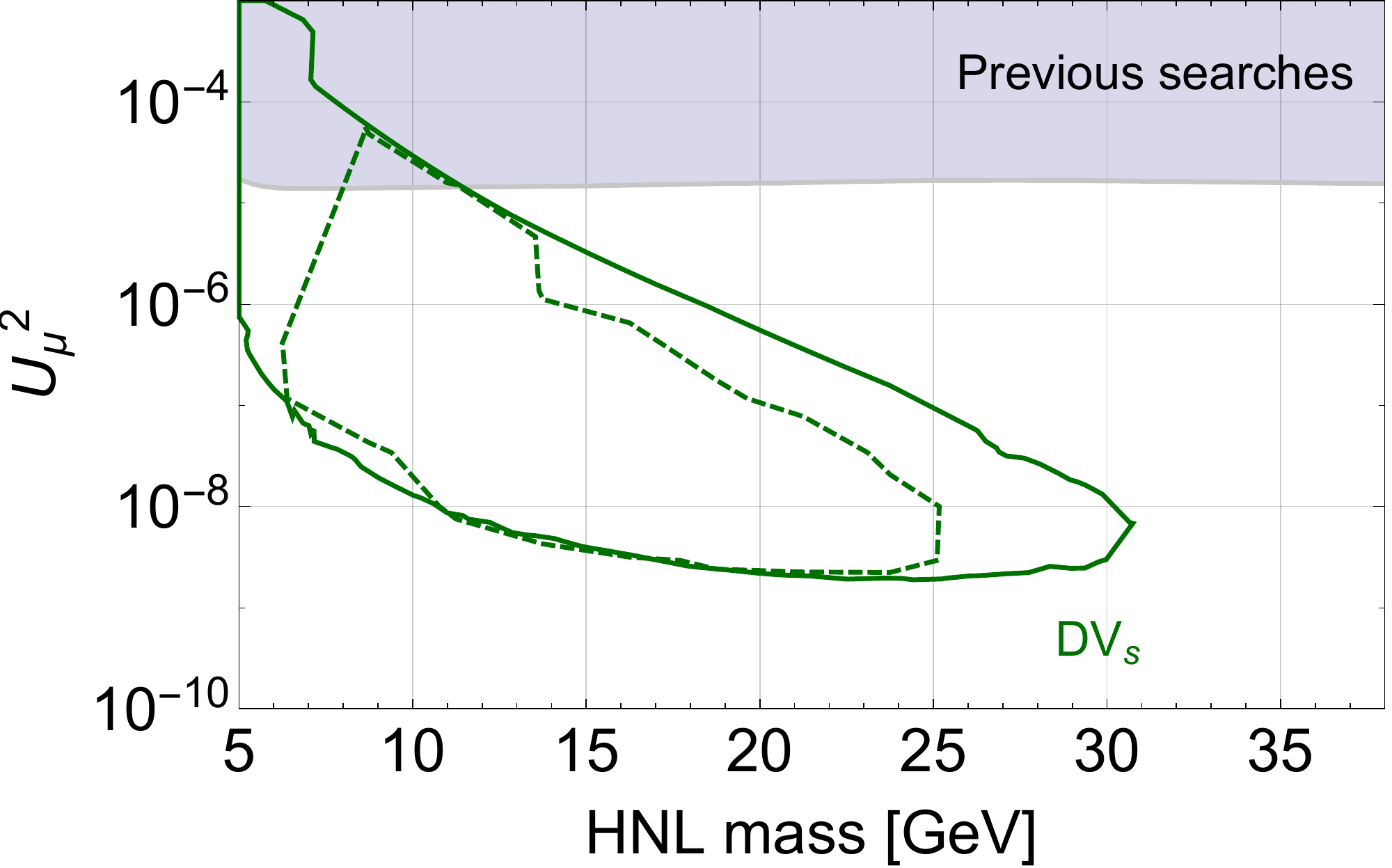}
    \\
    \includegraphics[width=0.45\textwidth,draft=false]{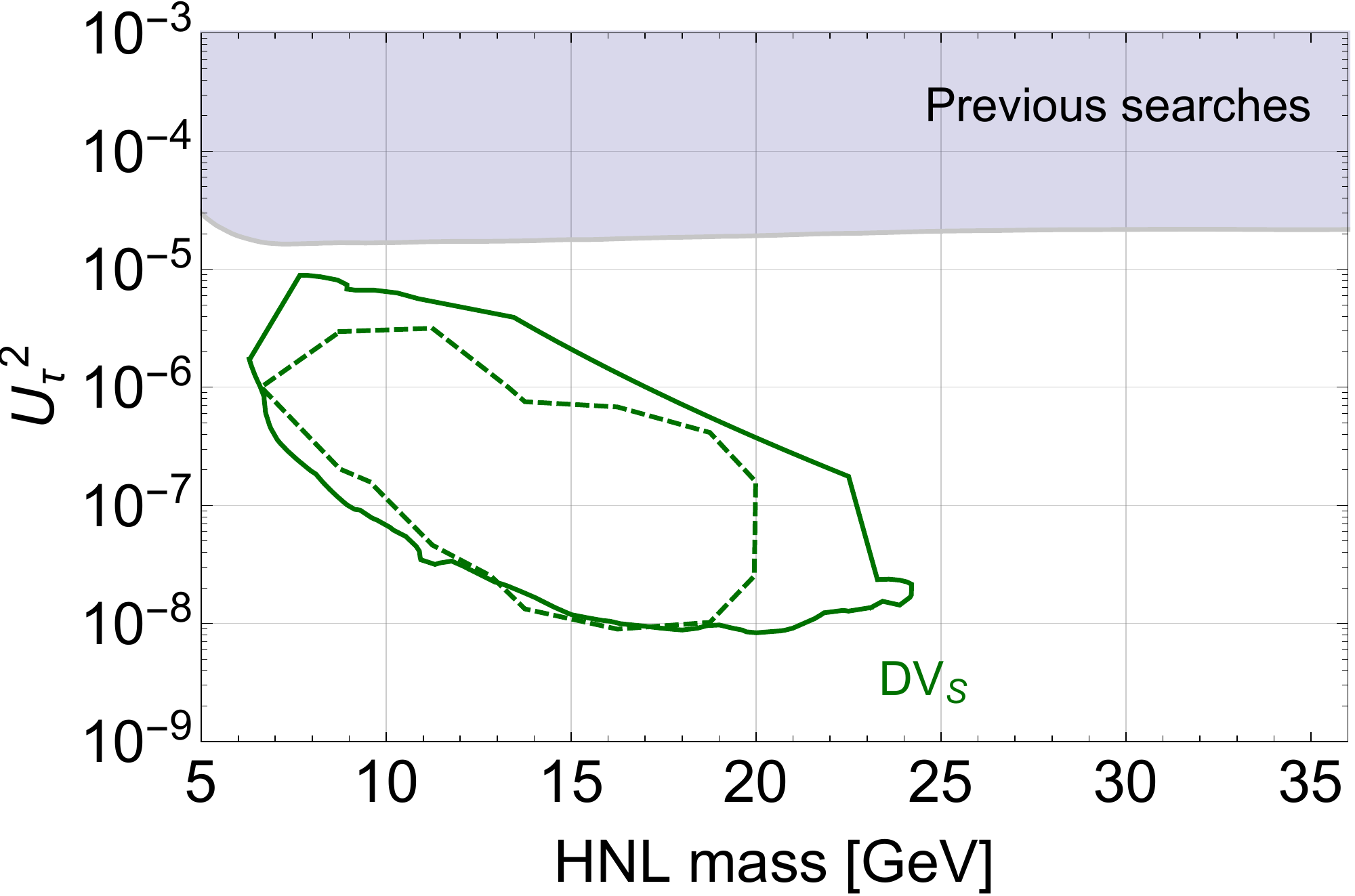}
    \caption{The sensitivity of DV searches with the ATLAS inner tracker
      (short DV, DV$_\mathrm{S}$).
      The green solid line shows our estimate, while the green dashed line
      shows the sensitivity based on the MC simulations from~\cite{Cottin:2018nms}.}
    \label{fig:short-dv-atlas}
\end{figure}

\begin{figure}
    \centering
    \includegraphics[width=0.45\textwidth]{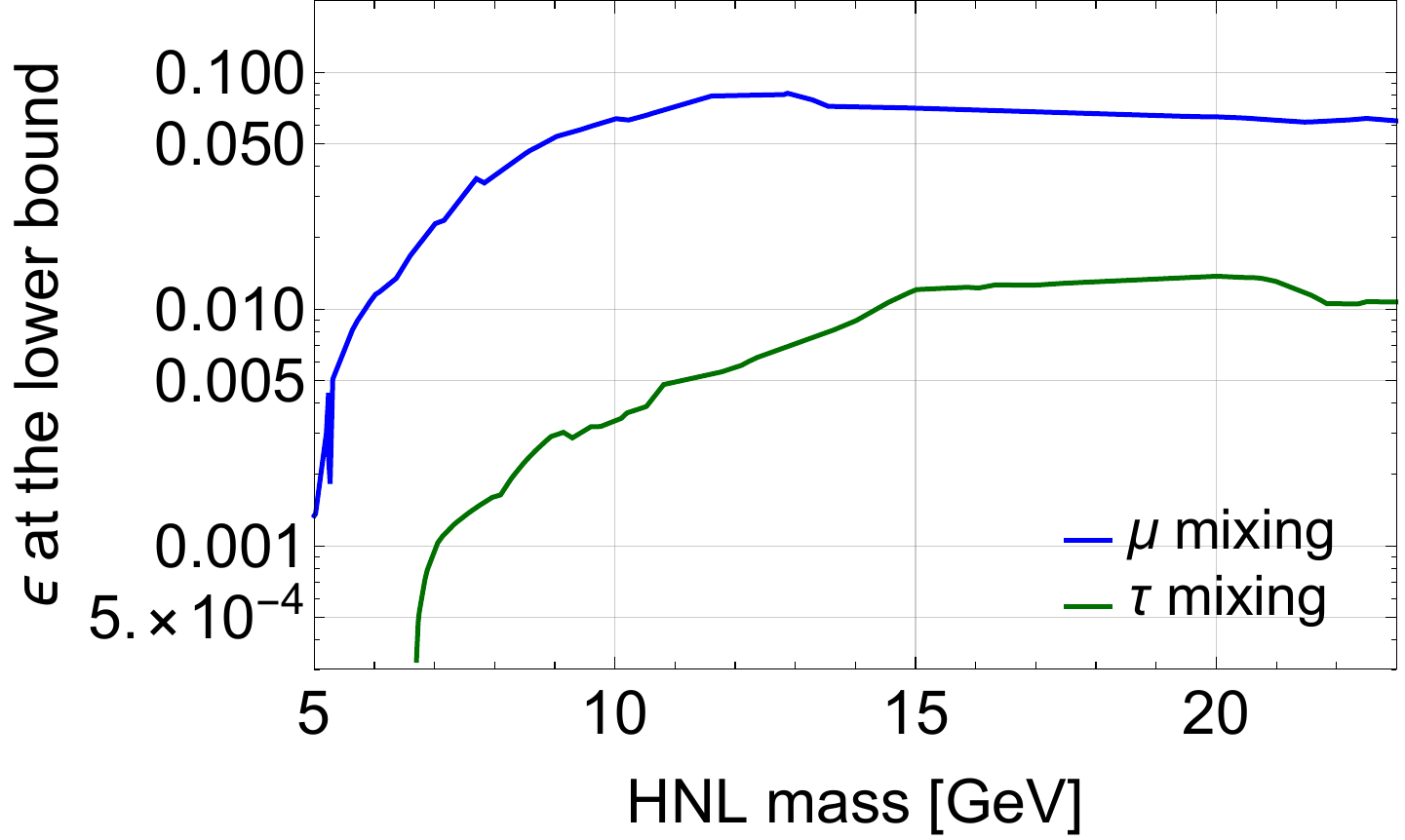}
    \caption{The efficiency for the ``short DV'' search schemes for mixing with $\nu_{\mu}$ (blue curve) and $\nu_\tau$ (green curve).}
    \label{fig:efficiency-dvs}
\end{figure}

Our resulting estimates of the sensitivity for ATLAS/CMS short DV searches, together with the sensitivity estimates from the simulations in~\cite{Cottin:2018nms}, are shown in Fig.~\ref{fig:short-dv-atlas}.
We limit our analysis to HNLs with $m_{N} > 5\text{ GeV}$. For lower masses production from $B$ mesons starts to contribute/dominate the production of HNL. For completeness, we show in Fig.~\ref{fig:efficiency-dvs} the value of the efficiencies for mixing  with $\nu_{\mu}$ and $\nu_\tau$ close to the lower boundary of the sensitivity region.

\begin{figure*}[t!]
    \centering
    \includegraphics[width=0.45\textwidth]{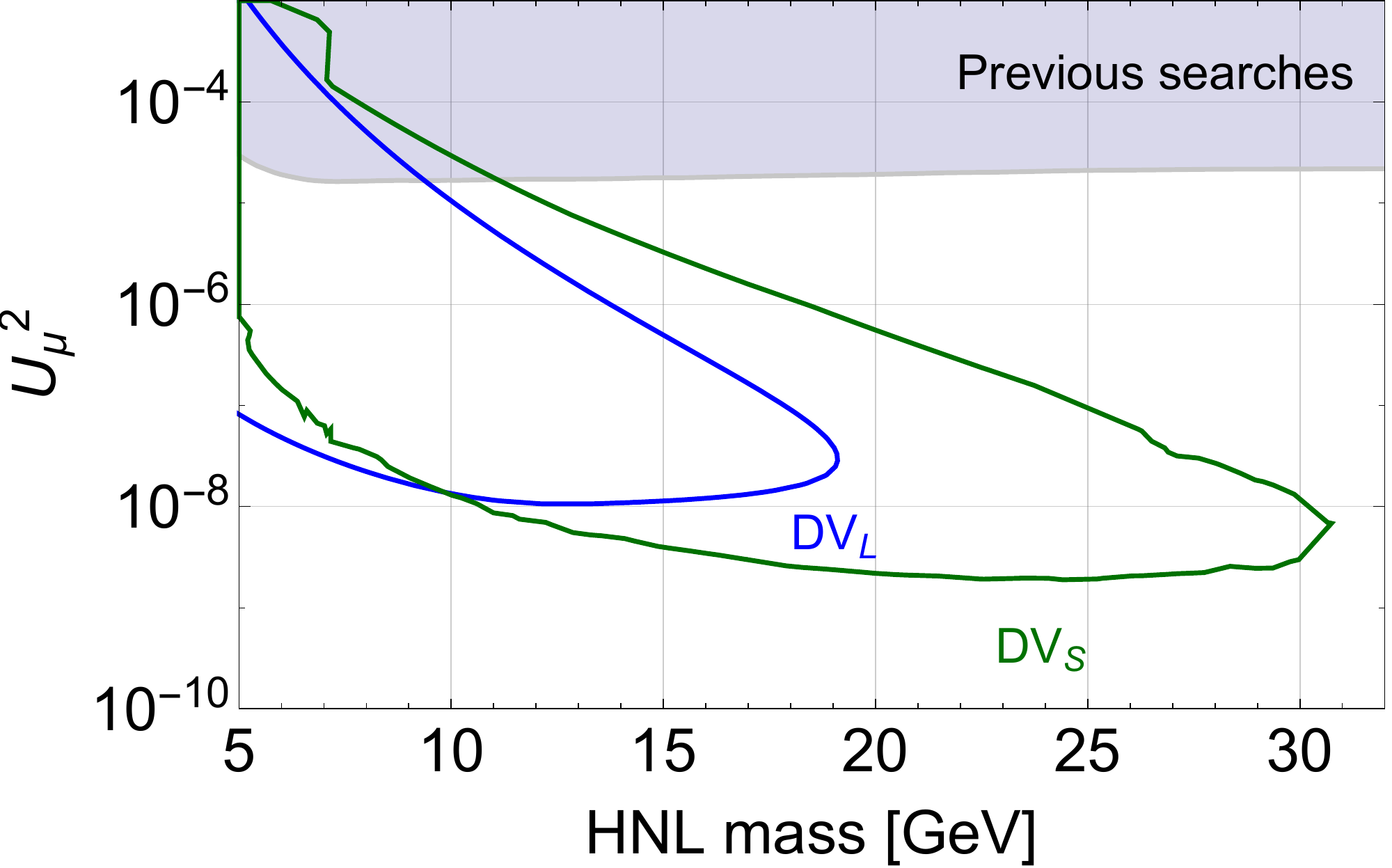}
    \includegraphics[width=0.45\textwidth]{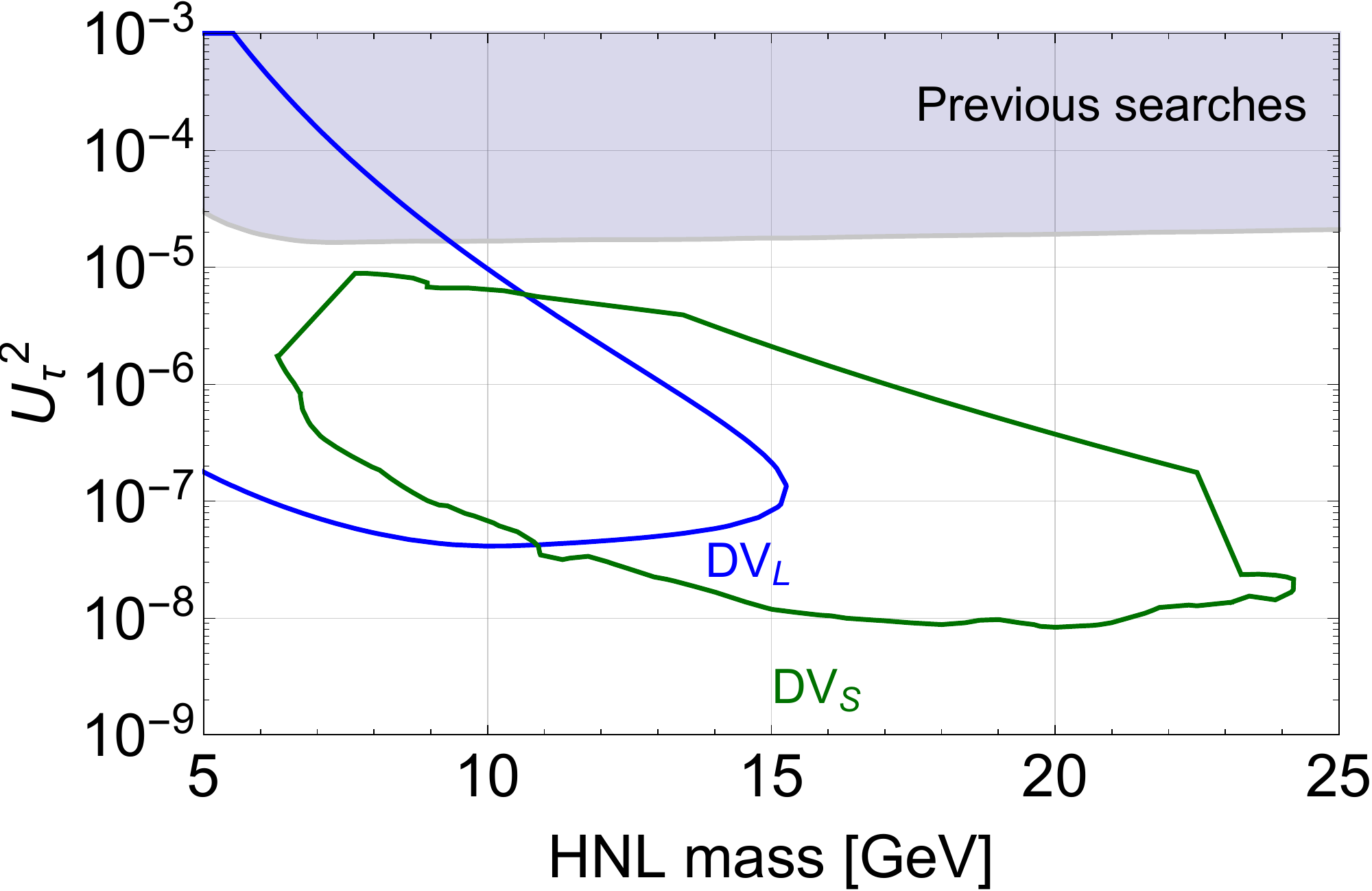}
    \caption{Dependence of the sensitivity of the ``long'' and ``short'' DV search schemes to the HNLs mixing with $\nu_{\mu}$ (left plot) and $\nu_{\tau}$ (right plot).}
    \label{fig:long-short-comparison}
\end{figure*}
\begin{figure*}[t!]
  \begin{minipage}{0.45\textwidth}
    \centering
    \includegraphics[width=\textwidth,draft=false]{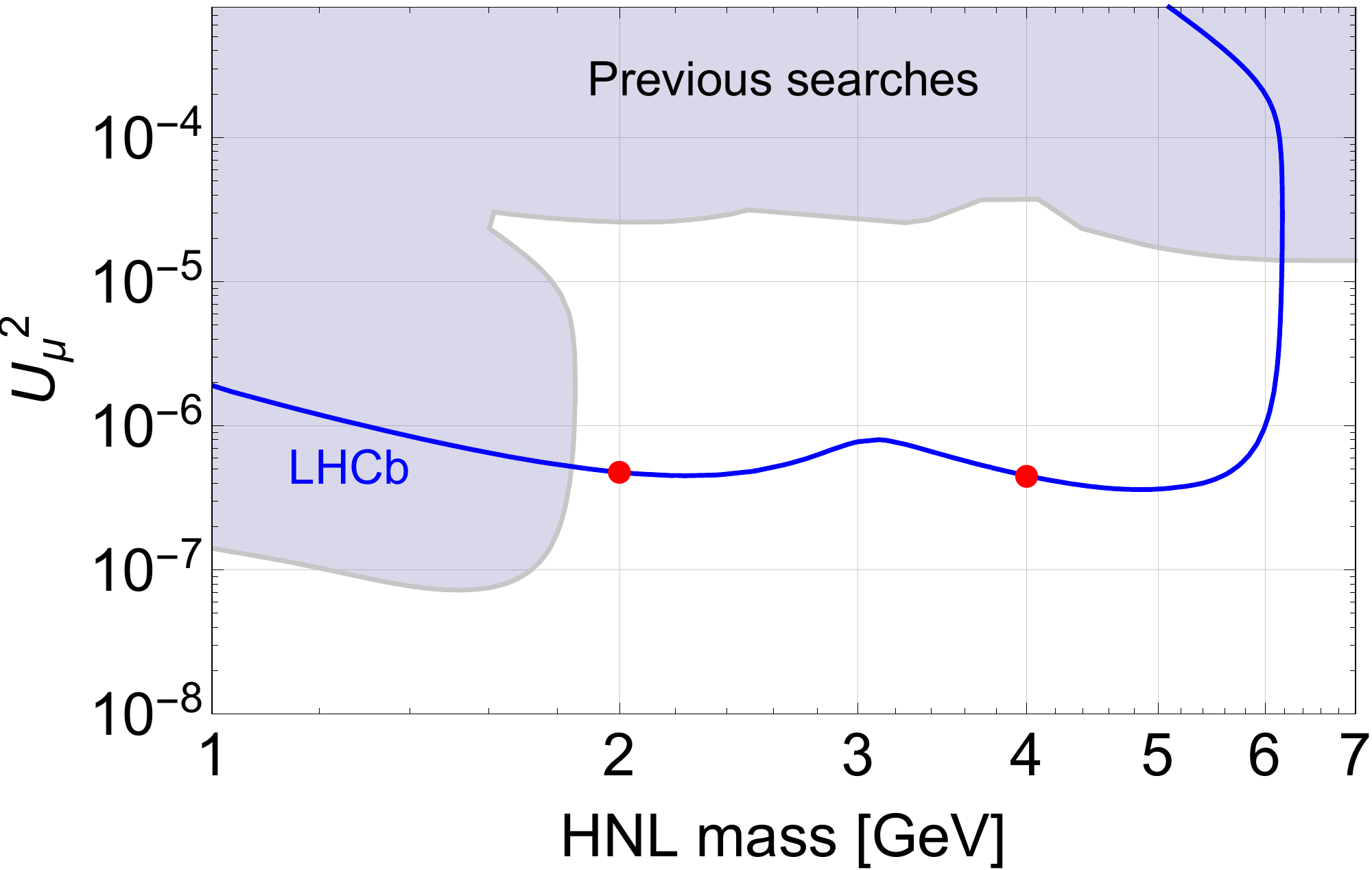}
    \end{minipage}~~\begin{minipage}{0.45\textwidth}
    \centering
    \includegraphics[width=\textwidth,draft=false]{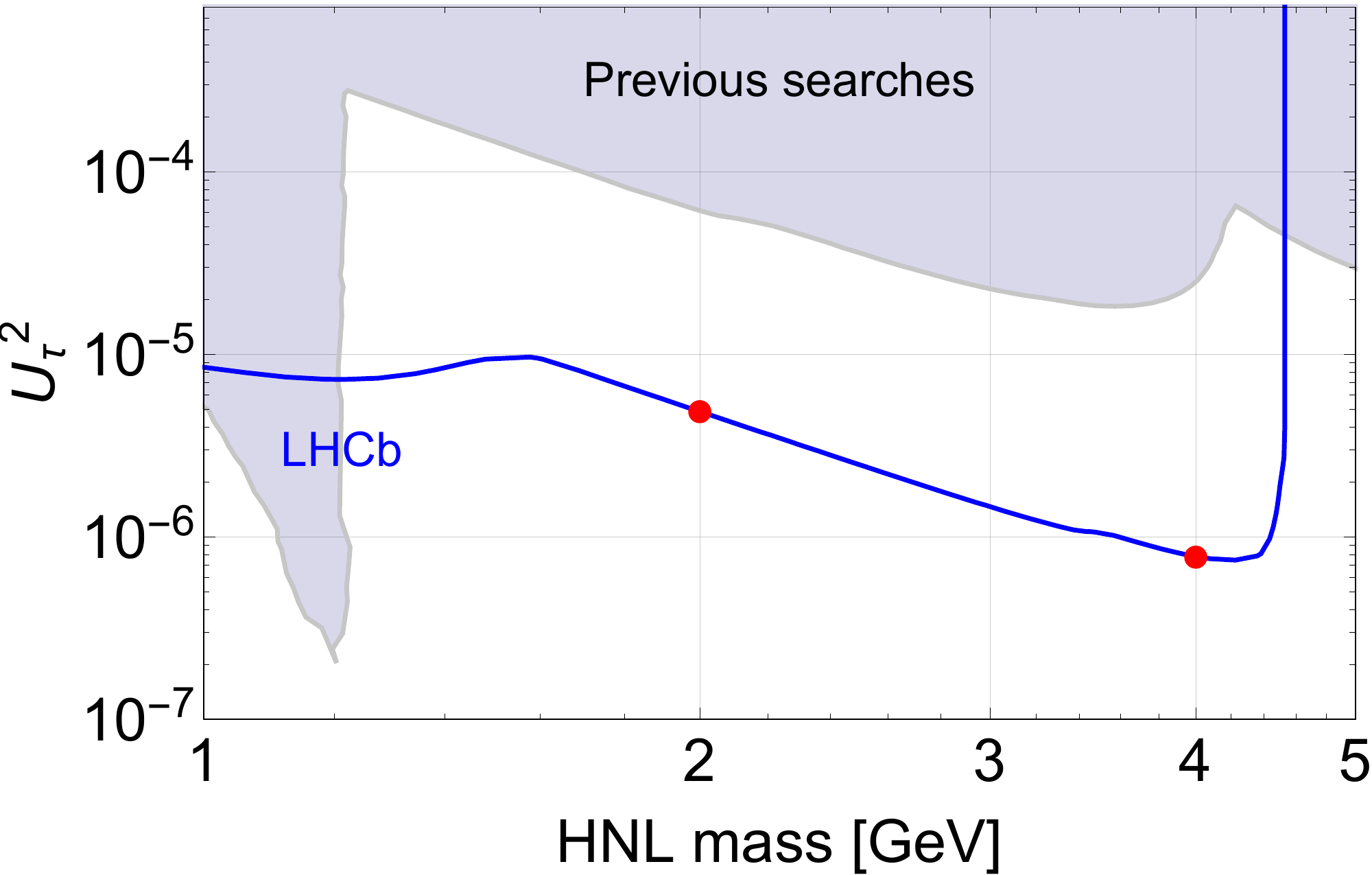}
    \end{minipage}
    \caption{The sensitivity of DV searches at LCHb in the high luminosity phase with $\mathcal L = \unit[300]{fb^{-1}}$. 
    Red points are the estimates of the lower bound~\protect\eqref{eq:lower-bound-lhcb} for particular masses from Table~\protect\ref{tab:lower-bound-lhcb}. 
    Bounds from the previous experiments are based on~\cite{Alekhin:2015byh}.}
    \label{fig:LHCb}
\end{figure*}
We expect that the DV searches with the CMS inner tracker will provide similar sensitivity.

\myparagraph{Possible alternatives to the short DV strategy.}
One of the selection criterion that strongly affects the efficiency is the requirement of at least four charged tracks needed to suppress hadronic background. An alternative way to suppress it is to search for DVs events with three leptons -- one prompt lepton originating from the decay $W \to N+l$ and two other from the decay $N \to l + l' +X$.
Let us estimate the sensitivities of this ``three lepton scheme'' for the HNLs
mixing with $\nu_{\mu}$ when all 3 leptons are muons.
For the efficiency of the three leptons scheme we have
\begin{equation}
\epsilon_{3\mu} = \epsilon_{\text{2 track DV}} \cdot \epsilon_{\mu\mu}\cdot \epsilon_{\text{prompt}}\cdot \text{Br}_{N\to \mu\mu \nu} \approx 3\cdot 10^{-3},
\end{equation}
where $\epsilon_{\text{2 track DV}} \approx 0.1$ is the efficiency of the reconstruction of the two-track DV, $\epsilon_{\mu\mu\nu}\approx 0.6$ is the efficiency of the reconstruction of two muons from the DV, and $\epsilon_{\text{prompt}} \approx 0.9$ is the efficiency of the reconstruction of the prompt muon originating from $W^\pm \to \mu^\pm +  N$.\footnote{We are grateful to 
Philippe Mermod for providing us with these efficiencies} The branching fraction $\text{Br}_{N \to \mu\mu \nu} \simeq 6\cdot 10^{-2}$~\cite{Bondarenko:2018ptm}.
We see that $\epsilon_{3\mu} $ is always smaller than the efficiency of the scheme from~\cite{Cottin:2018kmq,Cottin:2018nms} at the lower boundary of the sensitivity region, see Fig.~\ref{fig:efficiency-dvs}. Therefore we conclude that the scheme with two leptons is less efficient for the detection of the HNLs.

\myparagraph{Long DV at CMS.}
An alternative way to search for DV is to use the CMS muon tracker and to reconstruct the displaced vertex from two muons (see~\cite{Bondarenko:2019} for more details). This opens a possibility reconstructing a decay point at distances as far as $l_{\text{max}} \simeq 3\text{ m}$~\cite{Bayatian:2006nff}, that is why we call this method ``long DV'' scheme.
The selection criteria are:
\begin{compactitem}[--]
    \item Two displaced muon tracks, each with $p_{T} > 5\text{ GeV}$;
    \item The position of the DV must be within $l_{\text{min}} = 2\text{ cm}$ and $l_{\text{max}} = 3\text{ m}$.
\end{compactitem}
Both cuts are essential to make the long DV searches background free.
The comparison of the sensitivities of the ``short DV'' and ``long DV'' schemes is shown in Fig.~\ref{fig:long-short-comparison}.

\begin{figure*}[!t]
  \centering \includegraphics[width=0.5\textwidth,draft=false,clip=true, trim=0 25 0 0]{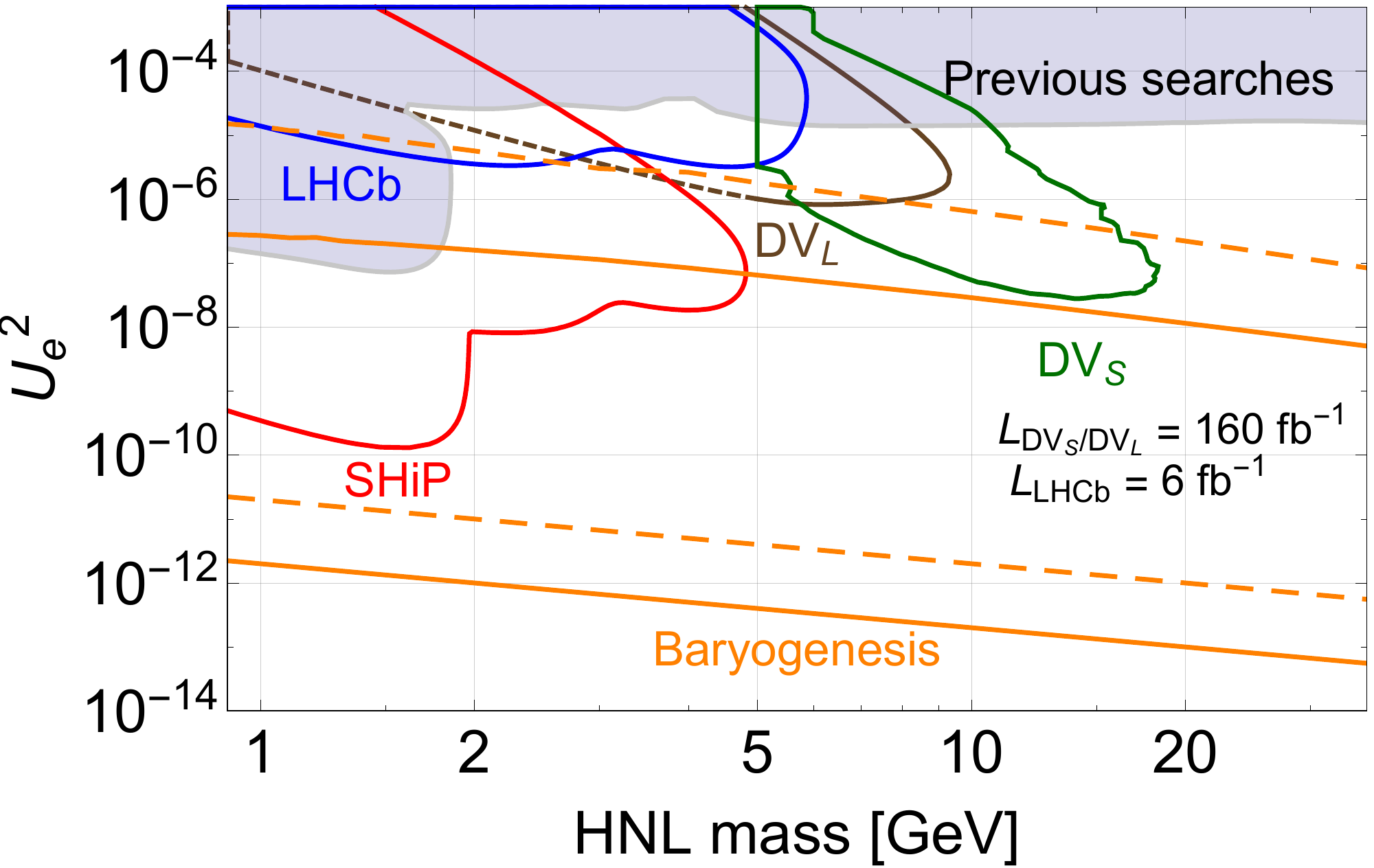}~
  \includegraphics[width=0.5\textwidth,draft=false,clip=true, trim=0 25 0 0]{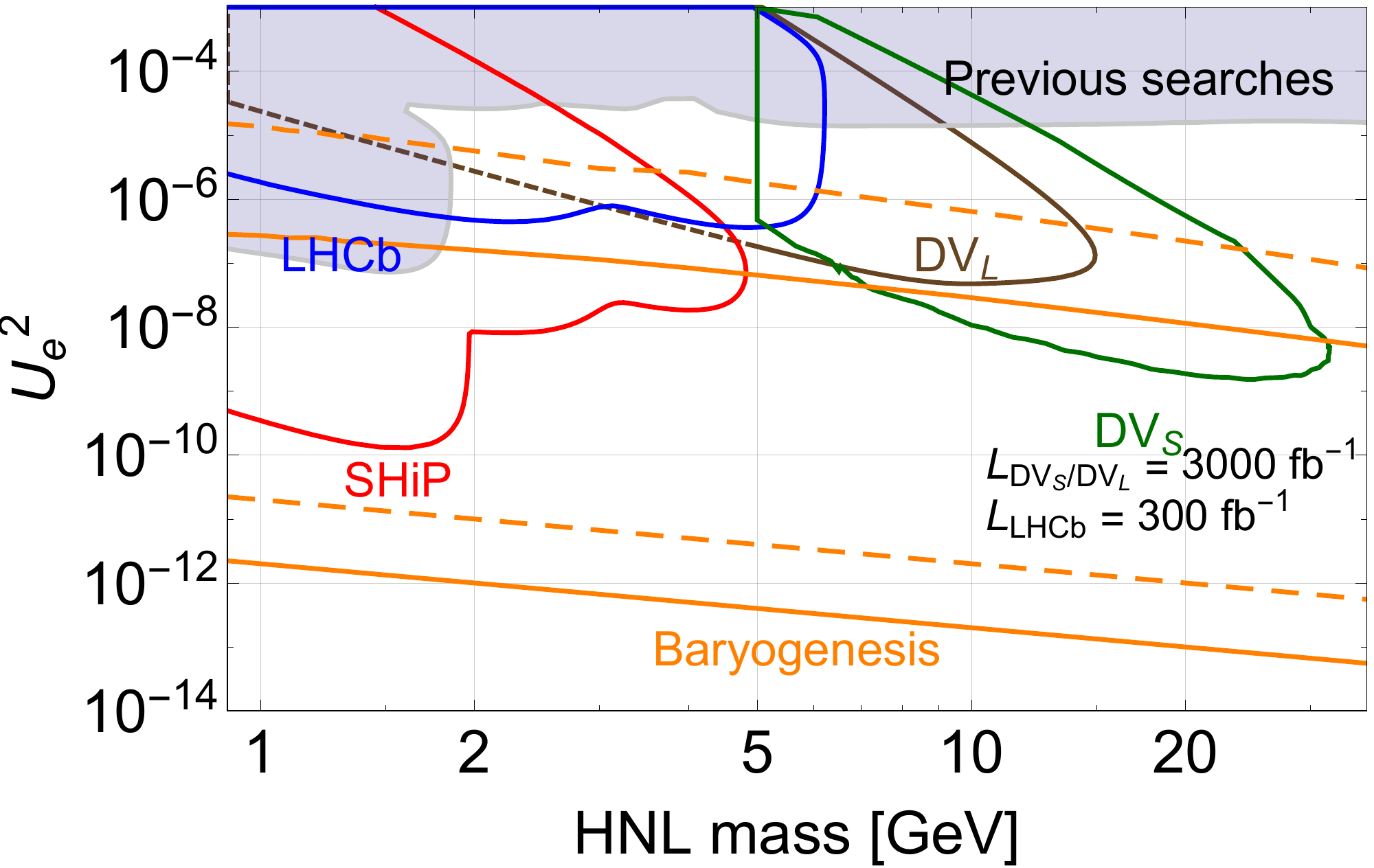}\\
  \includegraphics[width=0.5\textwidth,draft=false,clip=true, trim=0 25 0 0]{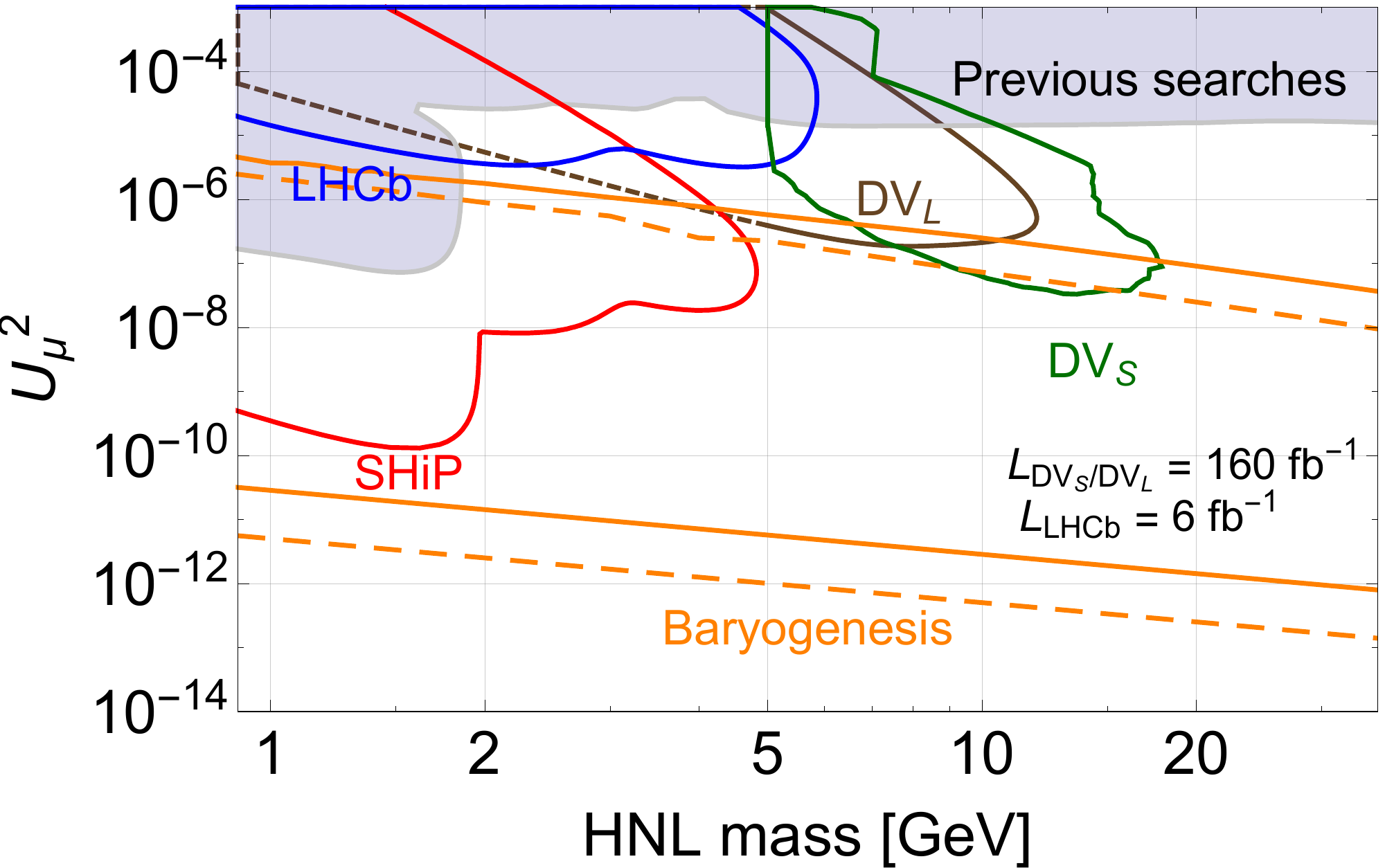}~
  \includegraphics[width=0.5\textwidth,draft=false,clip=true, trim=0 25 0 0]{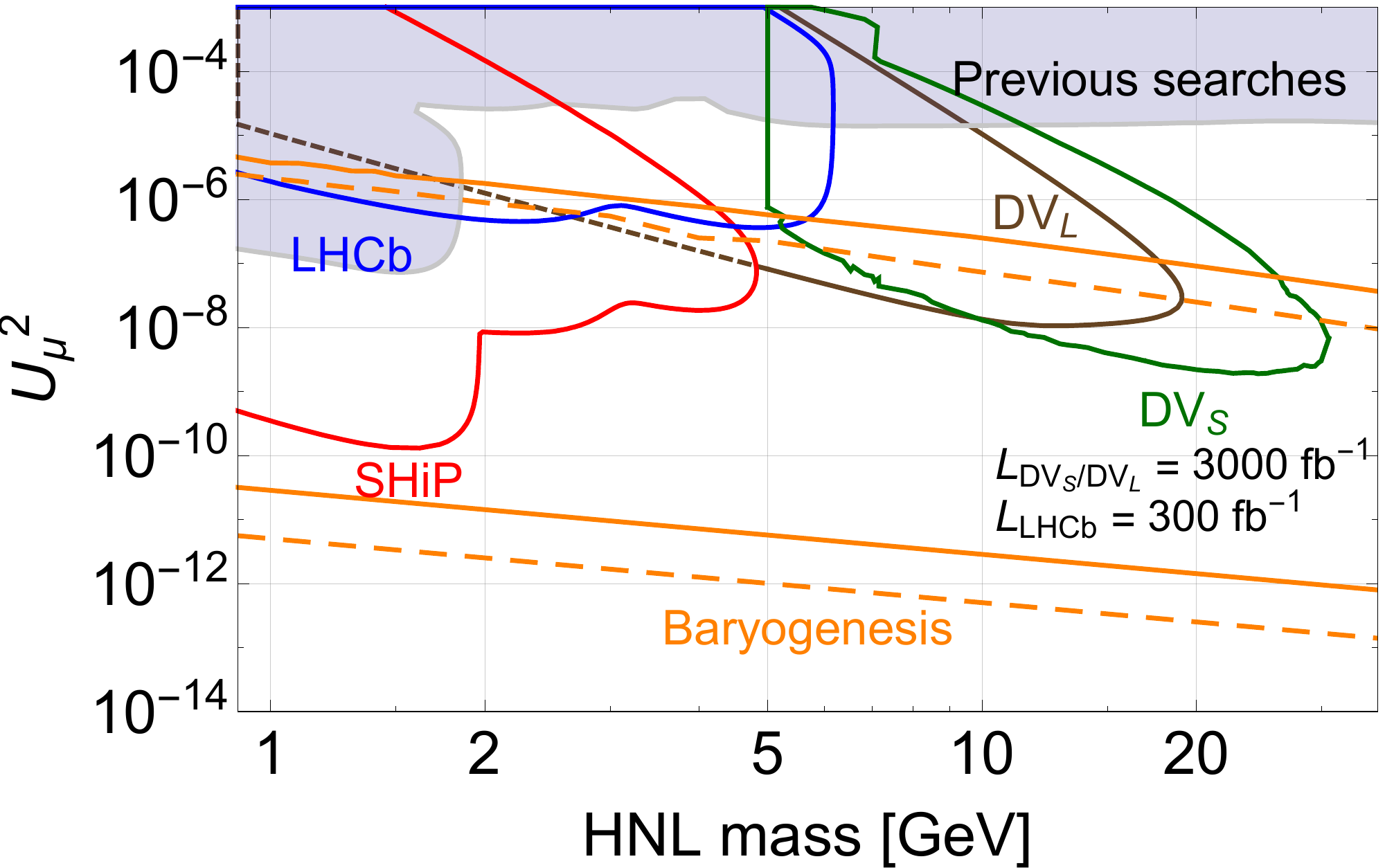}\\
  \includegraphics[width=0.5\textwidth,draft=false]{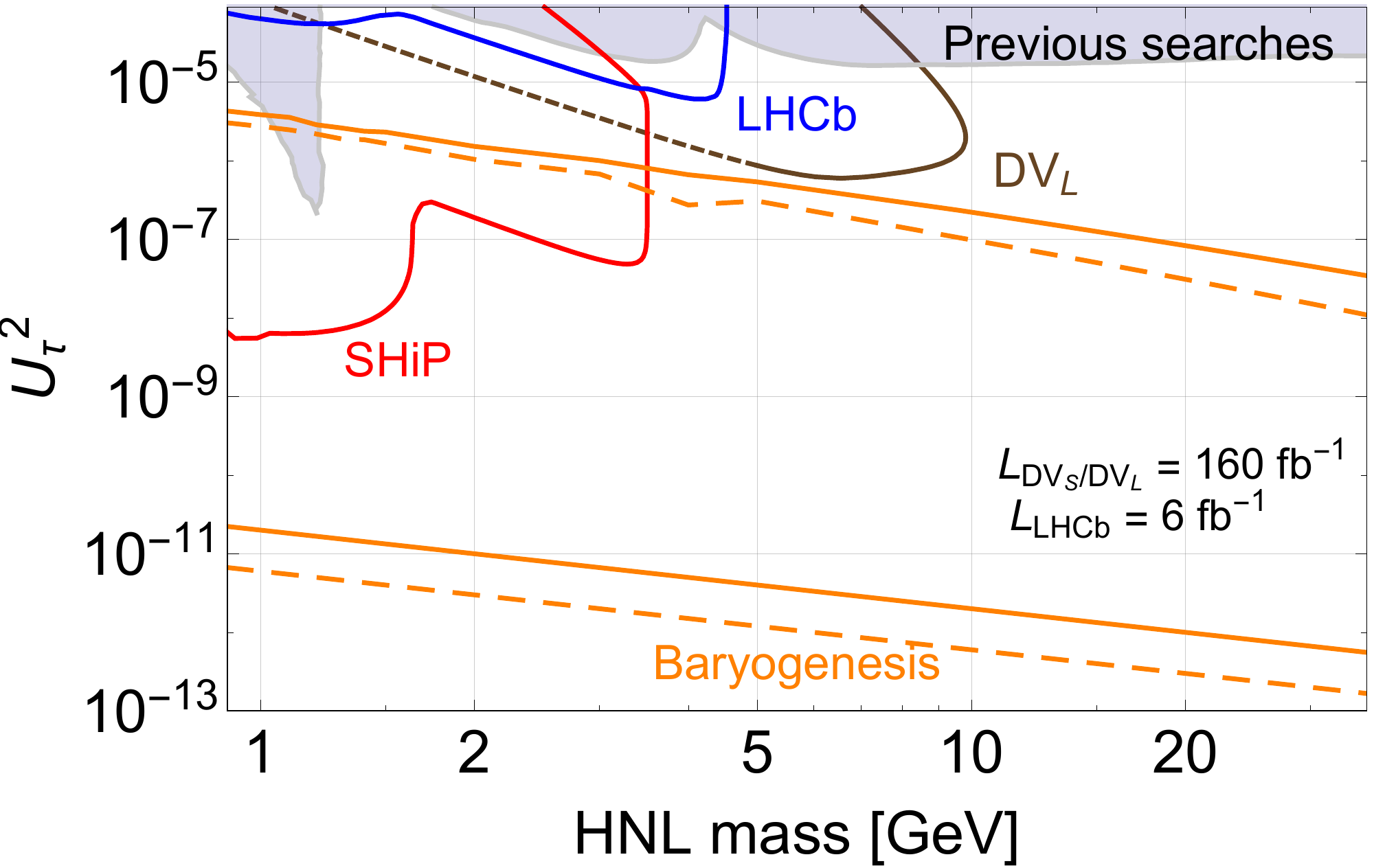}~
  \includegraphics[width=0.5\textwidth,draft=false]{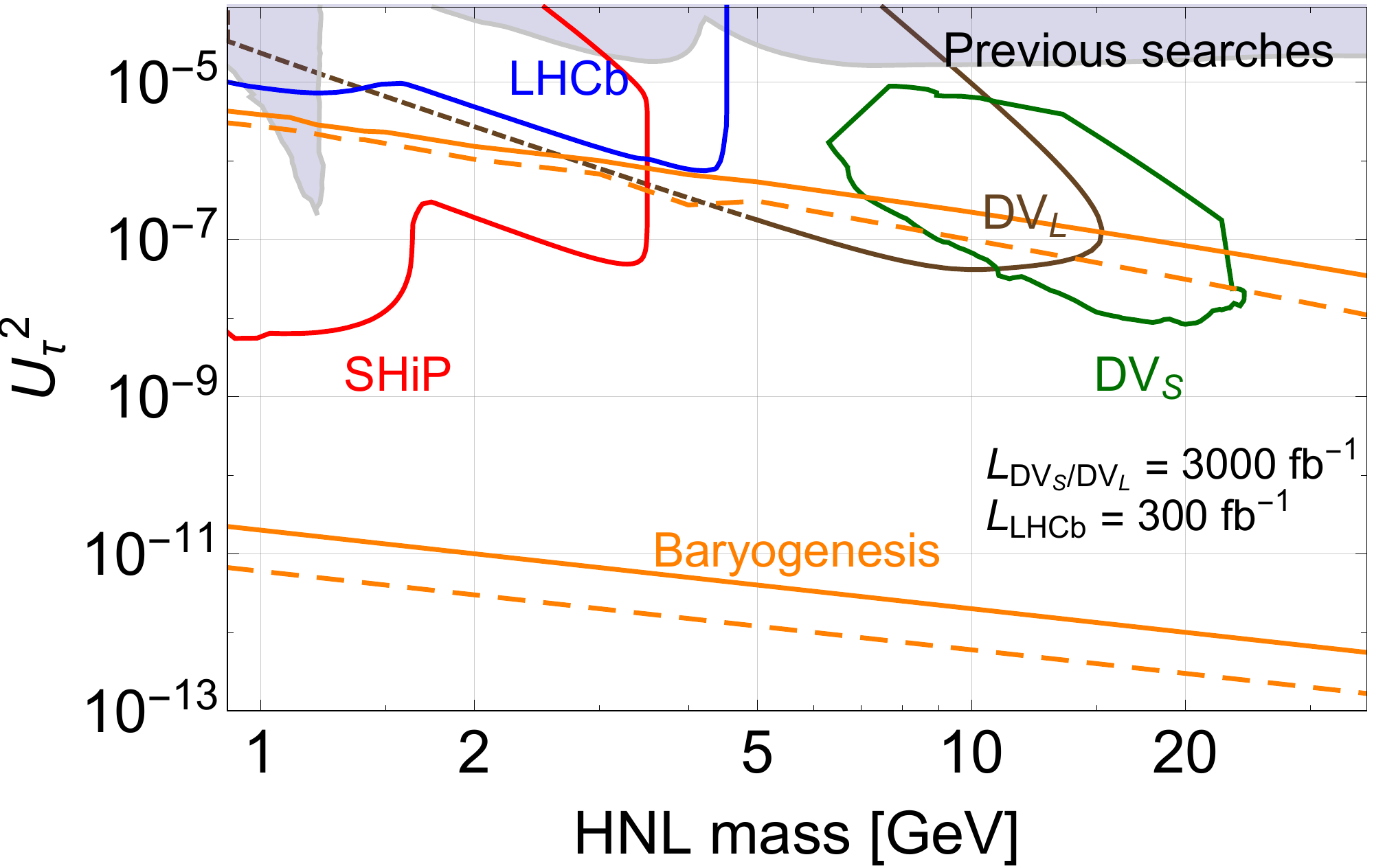}
  \caption{The parameter space of the HNLs in the $\nu$MSM. The observed BAU can be generated in the region between the orange lines. The sensitivity estimates from SHiP and DV searches at ATLAS (dark green, short DV), CMS (brown, long DV) and LHCb (blue) are shown. Short-dashed brown line below $5$~GeV corresponds to the estimate of  sensitivity of the long DV scheme under the assumption of zero background.
    \textit{Left panels} shows the estimates for current LHC luminosities; \textit{right panels} show the projections for the end of high-luminosity  phase. Solid orange lines are for the normal neutrino mass ordering, the dashed lines are for the inverted ordering. The orange lines are the combination of~\protect\cite{Eijima:2018qke} ($m_N \lesssim \unit[10]{GeV}$) and of this work (higher masses). Bounds from the previous experiments (shaded regions) are from~\cite{Alekhin:2015byh}.}
    \label{fig:models-probed-ue}
\end{figure*}

\myparagraph{DV at LHCb.}
\label{sec:lhcb}
We identify an HNL decay event at LHCb as a DV event if it passes the following selection criteria adapted from~\cite{Aaij:2014aba}:
\begin{compactitem}[--]
    \item Decay products must be produced in pseudorapidity range $2<\eta<5$.
    \item The single muon must have $p_{T} > 1.64\text{ GeV}$ to pass the trigger.
    \item Hadrons must have $p > 2\text{ GeV}$ and $p_{T} > 1.1\text{ GeV}$ to be tracked.
    \item Lepton products of HNL decay should have $p > 3\text{ GeV}$ and $p_T > 0.75\text{ GeV}$.
\end{compactitem}
Following~\cite{Aaij:2014aba}, we estimate the corresponding efficiency as
$\epsilon \sim 10^{-2}$ for all visible decay channels.

The main parameters for the LHCb experiment are given in Table~\ref{tab:parent}. We notice that at the energies of the LHC for large masses of the HNL ($m_{N} \simeq 3\text{ GeV}$ in the case of the mixing with $\nu_{e/\mu}$ and $m_{N} \simeq 2\text{ GeV}$ in the case of the mixing with $\nu_{\tau}$) the main production channel is the 2-body leptonic decays of the $B_{c}$ mesons (see, e.g.,~\cite{Bondarenko:2018ptm,SHiP:2018xqw}). This makes possible to probe HNL masses up to $m_{B_{c}}\approx 6.3\text{ GeV}$.
The mixing angle $U^{2}$ at the lower bound of the sensitivity is given by
\begin{equation}
U^{2}_{\text{lb}} = 2.6\cdot 10^{-6} \sqrt{\frac{300}{\mathcal{L} [\text{fb}^{-1}]}\frac{10^{-2}}{\epsilon}  \frac{c\tau_{N}}{1\text{ m}}   \frac{\langle\gamma_{N}\rangle}{\sum_{B}f_{b\to B}\text{Br}_{B\to N+X}}},
\label{eq:lower-bound-lhcb}
\end{equation}
The relevant parameters for masses of 2 and 4~GeV are summarized in Table~\ref{tab:lower-bound-lhcb}.

\begin{table}[h!]
  \centering
  \begin{tabular}{|c|c|c|c|c|c|}
    \hline
   Mixing & $m_{N}$, GeV & $f_{b\to B} \text{Br}_{B\to N+X}$ & $c\tau_{N}$, m & $\langle\gamma_{N}\rangle$  \\
    \hline
      $\mu$ & $2$ & $9.7 \cdot 10^{-3}$ & $1.7\cdot 10^{-5}$ & $22.5$   \\
       \hline
      $\mu$ & $4$ & $1.8\cdot 10^{-4}$ & $4.5\cdot 10^{-7}$ & $14.3$   \\
       \hline
      $\tau$  & $2$ & $2.4\cdot 10^{-4}$ & $4.5\cdot 10^{-5}$ & $22.5$ \\
       \hline
      $\tau$ & $4$ & $1.3\cdot 10^{-4}$ & $9.6\cdot 10^{-7}$  & $14.3$  \\
       \hline
  \end{tabular}
  \caption{The values of the parameters from the estimate~\eqref{eq:lower-bound-lhcb} of the lower bound of the LHCb experiment for particular masses. We use the value of the momentum of the $B$ meson from Sec.~\ref{tab:parent}, the HNL production branching from the Fig.~\ref{fig:branching} and the discussion above~\eqref{eq:lower-bound-lhcb} for the details.} 
  \label{tab:lower-bound-lhcb}
 \end{table}

The plot of the sensitivity is shown on Fig.~\ref{fig:LHCb}.

\section{Results}
\label{sec:results}
Our main results are given in Fig.~\ref{fig:models-probed-ue}. We show the sensitivities for two different luminosities: $\mathcal{L}_{\text{c}}$, which corresponds to the current accelerator statistics, and high luminosity $L_{\text{h}}$, which corresponds to LHC run 4. For LHCb they are $\mathcal{L}_{\text{c}} = 6.7\text{ fb}^{-1}$ and $\mathcal{L}_{\text{h}} = 300\text{ fb}^{-1}$, while for ATLAS/CMS they are $\mathcal{L}_{\text{c}} \approx 160\text{ fb}^{-1}$ and $\mathcal{L}_{\text{h}} = 3000\text{ fb}^{-1}$.

Our conclusions are the following: 
\begin{itemize}[--]
\item Intensity frontier experiments can probe the baryogenesis in the range $m_{N} \lesssim m_{B}$.
  For $m_{N}\lesssim m_{D}$ they are able to probe mixing angles almost all
  the way down to the lower boundary, determined by the requirement of the successful neutrino oscillations and BAU;
\item DV search schemes at ATLAS and CMS have a potential to probe the baryogenesis parameter space in the mass range $m_{N}\gtrsim m_{B}$ and are complementary to the intensity frontier experiments;
\item In the mass region $m_{N} \lesssim m_{B}$ there is a region that cannot be probed by any of the above searches. We argued that the LHCb DV search for HNLs produced in $B$ mesons decays and perhaps CMS long DV search have an ability to cover this domain.
\end{itemize}

\textit{In conclusion} the combination of displaced vertex searches at the ATLAS, CMS and LHCb together with  intensity frontier experiments can enter the cosmologically interesting part of the HNL parameter space and has the potential to discover particles responsible for both neutrino oscillations and baryon asymmetry of the Universe.

\subsubsection*{Acknowledgements.}
We thank G.~Cottin, J.-C.~Helo, R.~Jacobsson, P.~Mermod, M.~Shaposhnikov, L.~Shchutska for discussions and comments on the manuscript. 
This project has received funding from the European Research Council (ERC)
under the European Union's Horizon 2020 research and innovation programme (GA 694896) and from the Netherlands Science Foundation (NWO/OCW).

\appendix
\bigskip

\section{Production of the HNLs from $W$-bosons}
\label{sec:production}
We estimate the number of the HNLs flying in the direction of the ATLAS and CMS experiments in the following way:
\begin{equation}
    N_{\text{prod}} = \sigma_{W}\times \mathcal{L}\times \epsilon_{\text{geom}},
\end{equation}
where $\sigma_{W}$ is the total production cross-section of the $W$ bosons at the LHC (we use $\sigma_{W}\approx 193\text{ nb}$ from the experimental paper~\cite{Aad:2016naf}) and $\epsilon_{\text{geom}}$ is the geometric acceptance (i.e. the amount of the HNLs that are produced in the direction of the pseudorapidity range of ATLAS $|\eta| < 2.5$). The total value of the branching ratio of the $W$ boson decay to HNLs is shown in Fig.~\ref{fig:branching}.

We estimate the momentum spectrum of the $W$ bosons $dN_{W}/dp_{W}$ by performing the leading order simulation of the process $p + p \to W+X$ in MadGraph 5. Using it, we obtained $\epsilon_{\text{geom}}$ and the average HNL energy $\langle E_{N}\rangle$ by calculating the distribution of the HNLs in the energy and the angle $\theta_{N}$ between the direction of motion of the HNL and the $W$ bosons (whose momenta are found to be collinear to the proton beam direction in the simulations):
\begin{equation}
\label{eq:hnls-from-w}
    \frac{d^{2}N_{N}}{dE_{N}d\theta_{N}} = \int dp_{W}\frac{dN_{W}}{dp_{W}}\times \frac{d^{2}\text{Br}_{W\to N+l}}{d\theta_{N}dE_{N}}\times P(\theta_{N})
\end{equation}
Here 
\begin{multline}
\frac{d^{2}\text{Br}_{W\to N+l}}{d\theta_{N}dE_{N}} = \frac{1}{\Gamma_{W}}\frac{|\mathcal{M}_{W \to l + N}|^{2}}{8 \pi}\times \\ \times \delta(m_{N}^{2}+m_{W}^{2} -2E_{N}E_{W}+2p_{N} p_{W}\cos(\theta_{N}))
\end{multline}
is the differential production branching ratio, and $P(\theta_{N})$ is a projector which takes the unit value if $\theta_{N}$ lies inside the range $|\eta| < 2.5$ and zero otherwise. 

\begin{figure*}
    \centering
    \includegraphics[width=0.45\textwidth]{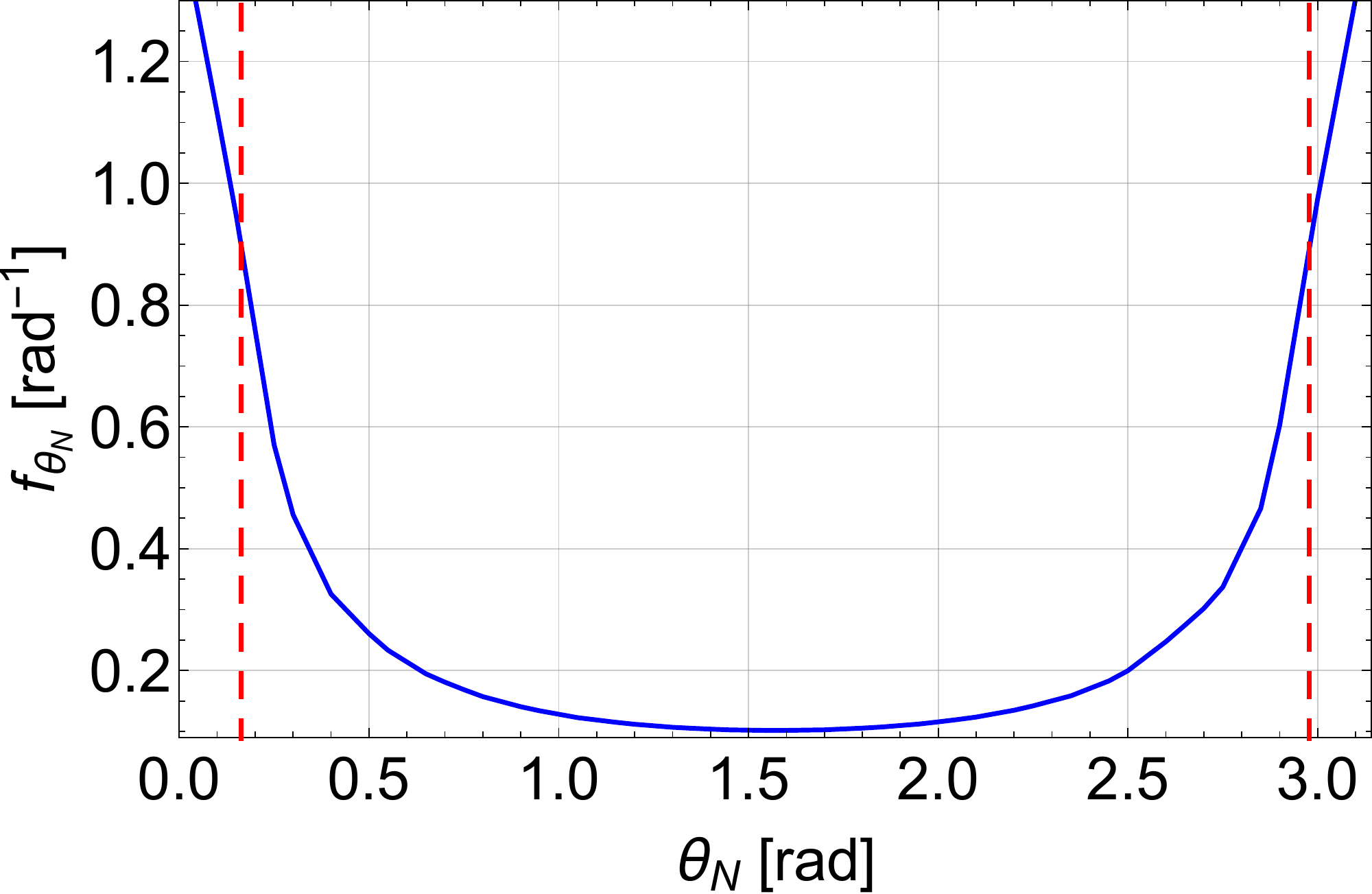}~
    \includegraphics[width=0.45\textwidth]{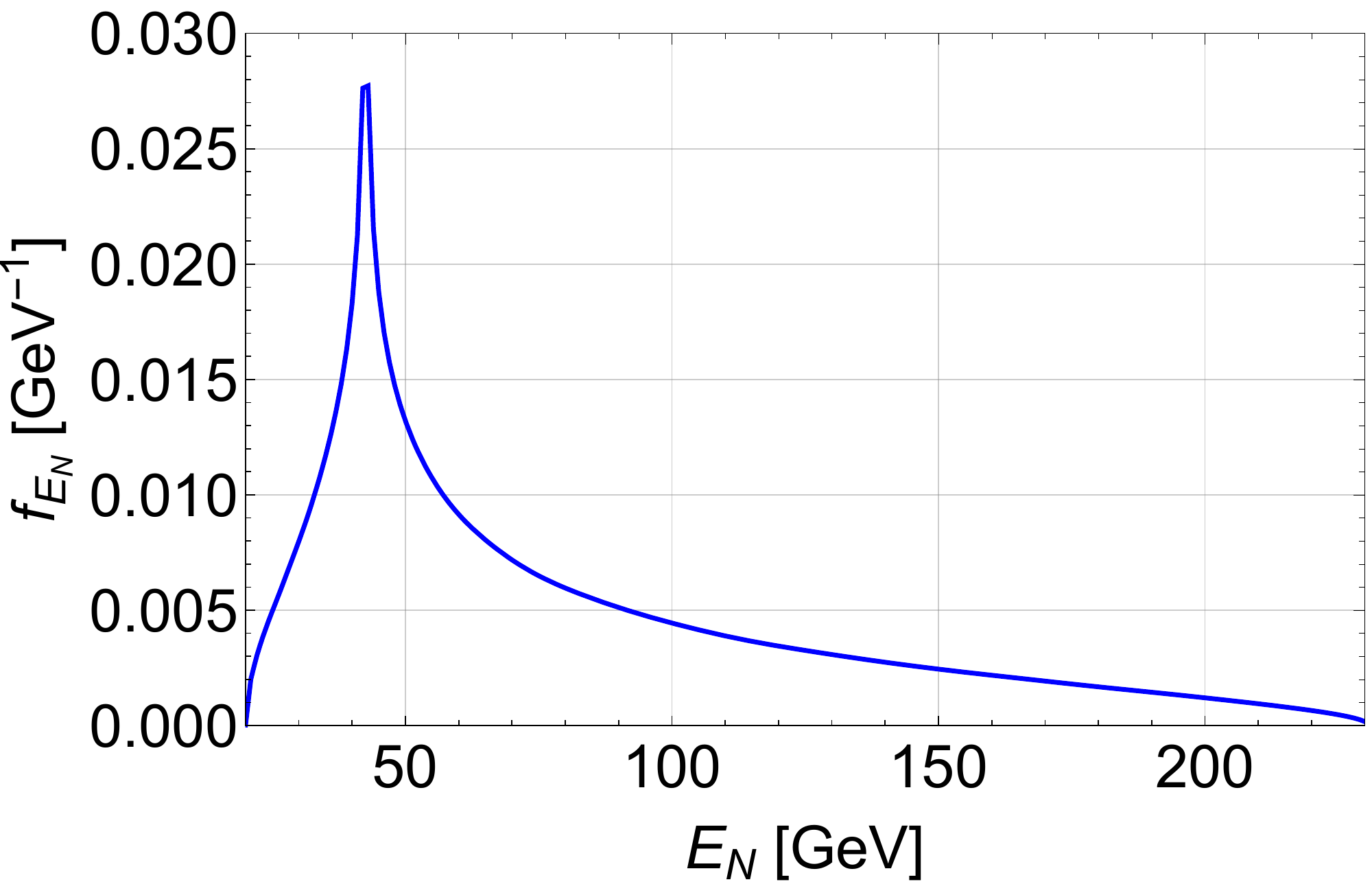}
    \caption{Left panel: the angular distribution of the HNLs produced in the decays of the $W$ bosons. The red dashed lines indicate the values of the angle corresponding to $\eta = \pm 2.5$. Right panel: the energy distribution of the HNLs flying in the direction of the ATLAS inner tracker. An HNL mass $m_{N} = 20\text{ GeV}$ is used. The peak around $E_{N} \approx M_{W}/2$ is caused by the contribution of the $W$ bosons produced with very low $p_{T}$, so that all of the HNLs have the same energy equal to the half of the $W$ boson mass.}
    \label{fig:hnl-distributions}
\end{figure*}

The resulting angular and  energy distributions of the HNLs at ATLAS are shown in Fig.~\ref{fig:hnl-distributions}. For the mass range $m_{N}\lesssim 30\text{ GeV}$ we obtained $\langle E_{N}\rangle \approx 80\text{ GeV}$ and $\epsilon_{\text{geom}} \approx 0.5$ with the accuracy $\sim 20\%$.

\bibliographystyle{JHEP}
\bibliography{ship}

\end{document}